\documentstyle[11pt,epsfig]{article}

\newcommand{\beq}{\begin{equation}}
\newcommand{\eeq}{\end{equation}}
\newcommand{\bea}{\begin{eqnarray}}
\newcommand{\eea}{\end{eqnarray}}

\def\laq{~\raise 0.4ex\hbox{$<$}\kern -0.8em\lower
0.62 ex\hbox{$\sim$}~}
\def\gaq{~\raise 0.4ex\hbox{$>$}\kern -0.7em\lower
0.62 ex\hbox{$\sim$}~}

\def \pa {\partial}

\def \ra {\rightarrow}

\def \Da {\Delta}
\def \b {\beta}
\def \a {\alpha}

\def \Ga {\Gamma}

\def \sg {\sigma}
\def \da {\delta}
\def \ep {\epsilon}
\def \r {\rho}
\def \om {\omega}

\begin{document}
\begin{titlepage}

\begin{flushright}
BA-TH/02-439\\
UNIL-IPT 02-5\\
CERN-TH/2002-125 \\
hep-ph/0206131
\end{flushright}

\vspace*{1cm}

\begin{center}
\Large{\bf  Assisting pre-big bang phenomenology \\
through short-lived axions}

\vspace*{1cm}

\large{ V. Bozza${}^{1,2}$,
M. Gasperini${}^{3,4}$,
M. Giovannini${}^{5}$
and G. Veneziano${}^{6}$}

\bigskip
\normalsize

{\sl $^{(1)}$  Dipartimento di Fisica
``E. R. Caianiello", Universit\`a di
Salerno, \\
Via S. Allende, 84081 Baronissi (SA), Italy\\
\vspace{0.3cm}
$^{(2)}$ INFN, Sezione di
Napoli, Gruppo Collegato di Salerno, Salerno, Italy} \\
\vspace{0.3cm}
{\sl $^{(3)}$ Dipartimento di Fisica,
Universit\`a di Bari, \\
Via G. Amendola 173, 70126 Bari, Italy\\
\vspace{0.3cm}
$^{(4)}$ INFN, Sezione di Bari, Bari, Italy} \\
\vspace{0.3cm}
{\sl $^{(5)}$ Institute of Theoretical Physics, University of
Lausanne,\\
 BSP-1015 Dorigny, Switzerland} \\
\vspace{0.3cm}
{\sl $^{(6)}$  Theoretical Physics Division, CERN,
CH-1211 Geneva 23, Switzerland}


\vspace*{5mm}

\begin{abstract}
We present the results of a detailed study of how isocurvature axion
fluctuations are converted into adiabatic metric perturbations through
axion decay, and discuss the constraints on the parameters of  pre-big
bang cosmology needed for consistency with present CMB-anisotropy
data. The large-scale normalization of temperature fluctuations has a
non-trivial dependence both on the mass and on the initial value of the
axion. In the simplest, minimal models of pre-big bang inflation,
consistency with the COBE normalization requires a slightly tilted
(blue) spectrum, while a strictly  scale-invariant spectrum requires mild
modifications of the minimal  backgrounds at large curvature and/or
string coupling. 
\end{abstract}

\end{center}
\end{titlepage}

It is well known that, in the framework of pre-big bang cosmology
(see \cite{11,GV02} for recent reviews),
the primordial spectrum of scalar  (and tensor)
metric perturbations  is characterized  by a steep positive slope 
\cite{1}. Since the high-frequency normalization  of the
spectrum is fixed by the ratio of the string to the Planck
mass, the amplitude of metric
fluctuations turns out to be strongly suppressed at large scales, and
thus  unable to account for the CMB anisotropies observed by COBE
\cite{2} and by other satellite experiments \cite{3} (unless one
accepts rather drastic modifications of pre-big bang kinematics, as
recently suggested in \cite{4}).

A possible solution to this problem could be provided, a priori, by
the fluctuations of another background field of string theory,
in particular of the so-called Kalb--Ramond axion $\sg$ (the dual of the
NS-NS two-form appearing in the dimensionally reduced string effective
action \cite{5}). As first pointed out in \cite{6}, axionic  quantum
fluctuations of the vacuum are  amplified by pre-big bang
inflation, yielding a final spectrum whose index $n_{\sg}$ can vary,
depending on the evolution of extra dimensions. The scale-invariant
value of $n_{\sg} = 1$ is attained, amusingly enough, for particularly
symmetric evolutions of the
nine spatial dimensions in which critical superstrings consistently
propagate.

Indeed, even if no axion potential is present in the post-big
bang era,  a (generally non-Gaussian) spectrum of temperature
anisotropies can be  induced by the fluctuations of the massless
\cite{7,8}  axion field, at second order, through the so-called ``seed"
mechanism \cite{seed}. The same is true for a massive light axion that
has not decayed yet \cite{9}. Unfortunately, while the model is capable
of reproducing the low-multipole COBE data \cite{2}, it  clearly appears
\cite{10} to be disfavoured  with respect to standard inflationary
models when it comes to fitting  data in the acoustic-peaks region
\cite{3}.

An interesting alternative possibility,
first suggested in \cite{11}, and recently
discussed in detail  (and not exclusively within
 a string cosmology framework) in \cite{12}-\cite{Riotto}, uses a
general mechanism originally pointed out in  \cite{13}.
It is based on two basic assumptions: i) the constant value of the
axion background after the pre-big bang phase is displaced from the
minimum (conventionally defined as $\sg =0$) of the non-perturbative 
potential $V(\sg)$ generated in the post-big bang epoch; ii)  the axion
potential  is strong enough to induce a phase of axion dominance before
its decay into radiation. Under these two (rather plausible) 
assumptions, the initially amplified isocurvature axion fluctuations  can
be converted, without appreciable change of the spectrum, into
adiabatic (and  Gaussian) scalar curvature perturbations  until the time
of horizon re-entry: these can then possibly produce the observed CMB
anisotropies.

Various  aspects of this new mechanism have  already been
discussed in \cite{12} for the string theory axion, and in
\cite{14}-\cite{16} (mostly in the context of conventional inflationary
models) for the  case of a generic scalar field, dubbed the ``curvaton" in
\cite{14} (see also \cite{Riotto} for a possible application of this
mechanism to the ekpyrotic scenario). Here, after providing an explicit
derivation and computation of the conversion of axion fluctuations into
scalar  curvature perturbations, we shall discuss the constraints
imposed by the CMB data, and its possible consistency with the
small-scale normalization and tilts typical of pre-big bang models. It
will be argued, in particular, that a strictly flat spectrum is only
compatible  with non-minimal models of pre-big bang inflation. A
detailed account of this work, including numerical checks of the analytic
arguments and estimates  given here, will be presented in a forthcoming
paper \cite{BGGV}.
                                  
The conversion of the axionic isocurvature modes
(amplified during the pre-big-bang phase) into adiabatic curvature
inhomogeneities takes place  in the post-big-bang phase, where we
assume the dilaton to be frozen and the axion to be displaced from the
minimum of its potential. The relaxation of the axionic field towards the
minimum of its potential is determined by the following evolution
equations (units where $16 \pi G= 1$ are used)
\bea
&&
R_\mu^\nu-{1\over 2}\da_\mu^\nu R= {1\over 2}T_\mu^\nu +{1\over
2}\pa_\mu \sg \pa^\nu \sg + {1\over 2}\da_\mu^\nu \left[ V-{1\over
2}(\nabla_\mu \sg)^2\right], \nonumber\\
 &&
\nabla_\mu\nabla^\mu \sg + {\pa V \over \pa \sg}=0,
\label{1}
\eea
where $T_{\mu\nu}$ is the stress tensor of the matter sources,
which we assume to be dominated by the radiation fluid. In the case of 
a conformally flat metric, $g_{\mu\nu}= a^2\eta_{\mu\nu}$,  the time
and space components of such equations, together with the axion
evolution equation, can be written (in  conformal time and in three
spatial dimensions) respectively as
\bea
&&
6 {\cal H}^2= a^2 \left(\r_r +\r_\sg \right), ~~~~~~~~~~~~
4 {\cal H}' + 2 {\cal H}^2 = -a^2 \left(p_r +p_\sg \right),
\label{2}\\
&&
\sg '' + 2  {\cal H} \sg ' + a^2 {\pa V \over \pa \sg}
=0,
\label{3}
\eea
where ${\cal H}=a'/a= d(\ln a)/d\eta$, $\r_r=3p_r$ is the energy
density of the radiation fluid, and
\beq
\r_\sg = {1\over 2 a^2} \sg'^2 + V(\sg), ~~~~~~~
p_\sg = {1\over 2 a^2} \sg'^2 - V(\sg).
\label{4}
\eeq
The combination of Eqs. (\ref{2}) and (\ref{3})
leads to the conservation equation for the radiation
fluid, i.e.  $\r_r' + 4 {\cal H} \r_r=0$.

While the background is  radiation-dominated, at least at the
onset of the post-big-bang phase, the initial large-scale 
inhomogeneities are dominated by the (isocurvature) perturbations
coming from the pre-big bang amplification of the quantum fluctuations
of the axion. In order to study the conversion of isocurvature into 
scalar curvature (adiabatic) modes, the background Eqs.
(\ref{2}) and  (\ref{3}) should be supplemented by the evolution equations
of the scalar inhomogeneities, following from the perturbation of the 
Einstein equations (\ref{1}).

Thanks to the absence of anisotropic stresses, the
$i \not=j$ components of the perturbed Einstein equations imply
that the scalar metric fluctuations can be parametrized
in terms of a single gauge-invariant variable, the Bardeen potential
$\Phi$ \cite{17}. The full system of perturbed Einstein equations can
then be written as
\bea
&&
\Phi' + {\cal H} \Phi = {1\over 4} \chi \sg' +{1\over 3} a^2 \r_r v_r,
\label{6}\\
&&
\nabla^2 \Phi - 3 {\cal H} \left( \Phi' + {\cal H} \Phi\right)={1\over
4}a^2
\left(\r_r \da_r + \r_\sg \da _\sg\right),
\label{7}\\
&&
\Phi'' +3 {\cal H} \Phi' +\left({2\cal H}' + {\cal H}^2\right) \Phi =
{1\over 4} a^2\left({1\over 3} \r_r \da_r + \da p _\sg\right),
\label{8}\\
&&
\chi'' + 2 {\cal H} \chi' - \nabla^2\chi + a^2 {\pa^2 V \over \pa \sg^2}
\chi=4\sg' \Phi' -2 a^2{\pa V\over \pa \sg}\Phi,
 \label{9}
\eea
where the gauge-invariant variables $\chi$, $\delta\rho_{r}$, $v_{r}$
are, respectively, the axion, radiation density and velocity potential
fluctuations [with our conventions, in the longitudinal gauge the
velocity potential is defined by $\da T_i^0=(\r_r+p_r)\pa_iv_r$], and
where the following  variables 
\bea &&
\da_r= \da \r_r /\r_r, ~~~~~~~~~~~
\da_\sg= \da \r_\sg/\r_\sg, \nonumber \\
&&
\da \r_\sg  = - \Phi \left(\r_\sg + p_\sg\right)+ {\sg ' \chi'\over a^2}
+{\pa V \over \pa \sg} \chi,
\nonumber\\
&&
\da p_\sg  = - \Phi \left(\r_\sg + p_\sg\right)+ {\sg ' \chi'\over a^2}
-{\pa V \over \pa \sg} \chi
\label{10}
\eea
have been defined (we have also assumed $\da p_r=\da \r_r/3$). By
using the above perturbation
equations, together with the background relations (\ref{2}) and 
(\ref{3}), two useful equations for the evolution of the radiation
density
contrast and of the velocity potential can be finally obtained:
\beq
\da_r' = 4 \Phi' +{4\over 3} \nabla^2 v_r, ~~~~~~~~~~~~
v_r' ={1\over 4} \da_r + \Phi.
\label{12}
\eeq

We now suppose to start at $t=t_i$ with a radiation-dominated
phase in which the homogeneous axion background is initially constant
and non-vanishing, $\sg (t_i)=\sg_i \not=0$, $\sg'(t_i)=0$, providing a
subdominant (potential) energy density, $\r_\sg (t_i) =-p_\sg
(t_i)=V_i$ $ \ll H_i^2 \sim \r_r (t_i)$. The initial
conditions of Eqs. (\ref{6})--(\ref{9}) are imposed by assuming a given
spectrum of isocurvature axion fluctuations, $\chi_k(t_i)\not=0$, and 
a total absence of perturbations for the metric and the radiation fluid,
$\Phi(t_i)= \da_r(t_i)= v_r(t_i)=0$. The initial values  of the first
derivatives of the perturbation variables  are then fixed by enforcing
the momentum and Hamiltonian constraints, i.e. Eqs.  (\ref{6}) and
(\ref{7}).

Before discussing the origin of curvature fluctuations we
must  specify the details of the background
evolution. The axion, initially constant and subdominant, starts
oscillating at a curvature scale $ H_{\rm osc} \sim m$ (as can be
argued from Eq. (\ref{3})), and eventually decays (with gravitational
strength) into radiation,  at a scale $H_d \sim m^3/M_{\rm P}^2< H_{\rm
osc}$ (a process that must occur early enough, not to disturb the
subsequent standard evolution). When the axion is constant,
$\r_\sg$ behaves
like an effective cosmological constant, while during the oscillatory
phase its kinetic and potential energy density are equal on the
average, so that $\langle p_\sg \rangle=0$, and $\langle\r_\sg \rangle
\sim a^{-3}$ behaves like dust matter. Thus the radiation energy is
always diluted faster, $\r_r \sim a^{-4}$, and the axion background
tends to become dominant at a scale $H_\sg(t) \sim \sqrt{ V [\sg(t)]}$.

For an efficient conversion of the initial $\chi$ and
$\da_\sg$ fluctuations into $\Phi$ and $\da_r$ fluctuations
it is further required \cite{12}-\cite{15}, as we shall see, that the
decay occur after the beginning of the axion-dominated phase, i.e. when
$H_\sg > H_d$. Depending upon the relative values of $H_{\sigma}$ and
$H_{\rm osc}$ (i.e. depending upon the value of $\sigma_{i}$ in Planck
units) we have two different options which will  now be discussed
 separately. In order to perform explicit analytical estimates, we shall
assume here that $V(\sg)$ can be approximated by the quadratic form 
$m^2 \sg^2/2$. This is certainly true for $\sg_i \ll 1$, but it may be
expected to be a realistic approximation also for the range of values of
$\sg_i$ not much larger than $1$ (which, as we shall see, is the
appropriate range for a normalization of the spectrum compatible with
present data). Actually, for the periodic potential  expected for an axion
the  value of  $|\sg_i|$ is effectively bounded from above \cite{12}. 

$(1)$ If $\sg_i < 1$  then $H_\sg < H_{\rm osc}$, and the axion starts
oscillating  (at a scale $H \sim m$) when the Universe is still
radiation-dominated. During the oscillations the average potential
energy density decreases like $a^{-3}$, i.e. the typical amplitude of
oscillation decreases, following an $a^{-3/2}$ law, from its initial value
$\sg_i$  to the value $\sg_{\rm dom}$ at which $H= H_\sg  \sim m
\sg_{\rm dom}$. During this period $a \sim H^{-1/2}$ (as the background
is radiation-dominated), so that $\sg_{\rm dom} \sim \sg_i^4$, and
$H_\sg \sim m \sg_i^4$. Finally, the background remains axion-dominated
until the decay scale $H_d \sim m^3/M_{\rm P}^2$. This model of
background is thus consistent for  $H_i > H_{\rm osc}>H_\sg>H_d$,
namely for 
\beq 
1>\sg_i > (m/M_{\rm P})^{1/2},
\label{13}
\eeq
which allows for a wide range for $\sg_i$, if we recall the
cosmological bounds on the mass following from the decay of a
gravitationally coupled scalar \cite{18} (typically, $m>10$ TeV to
avoid disturbing standard nucleosynthesis).

$(2)$ If $\sg_i > 1$, and then $H_\sg >H_{\rm osc}$, the axion
starts dominating at the scale $H_\sg \sim m \sg_i$, which marks the
beginning of a phase of slow-roll inflation, lasting until the
curvature drops below the oscillation scale $H_{\rm osc} \sim m$.
Such a model of background is consistent for $H_\sg <H_1$, namely
for
\beq
{H_1/ m} >\sg_i >1,
\label{18}
\eeq
where $H_1$ (fixed around the string scale) corresponds to the
beginning of the radiation-dominated, post-big bang evolution.
During the inflationary phase the slow decrease of the Hubble scale
can be approximated (according to the background equations
(\ref{2}) and (\ref{3}))  by  $H(t) =\a m\sg_i -\b m^2 (t-t_\sg)$, where
$\a$ and $\b$
are dimensionless coefficients of order $1$. Inflation thus begins at
the epoch $t=t_\sg \sim 1/m\sg_i$, and lasts until the epoch $t =t_m
\sim (\sg_i-1)/m \sim \sg_i/m$.

Finally, if $\sg_i \sim 1$,  $H_\sg \sim H_{\rm osc}\sim m$, and the
beginning of the oscillating and of the axion-dominated phase are nearly
simultaneous. Let us now estimate, for these classes of backgrounds,
the evolution of the Bardeen potential generated by the primordial
axion fluctuations. 

It is convenient, for this purpose, to introduce  the 
gauge-invariant variable $\zeta$ representing the
spatial curvature  perturbation on uniform density (or
equivalently, at large scales, on comoving) hypersurfaces.
 For purely adiabatic
perturbations $\zeta$ is conserved (outside the horizon), 
and  can be written for a general background as
\cite{17}:
\beq
\zeta= -\Phi -{{\cal H} \Phi' + {\cal H}^2 \Phi
\over {\cal H}^2 - {\cal H}'}.
\label{21}
\eeq

Outside the horizon, Eq. (\ref{12}) gives  $4 \Phi = \da_r$;
 the sum of
the two background equations (\ref{2}) for the denominator
${\cal H}^2 - {\cal H}'$ and the Hamiltonian constraint
(\ref{7}) for the numerator ${\cal H} \Phi' + {\cal H}^2 \Phi$, allow 
 $\zeta$ to be rewritten in the convenient form
\beq
\zeta_k= {\r_\sg \da_\sg(k) -(3/4) ( \r_\sg + p_\sg) \da_r
(k) \over 4 \r_r + 3 ( \r_\sg + p_\sg)}.
\label{22}
\eeq
This expression has been obtained by neglecting the contribution of
spatial gradients in Eqs. (\ref{6})--(\ref{9}).  Numerical integration
shows \cite{BGGV} that the corrections coming from these terms are
indeed negligible for the large-scale modes leading to  the anisotropies
in the CMB.

Consider now the beginning of the post-big bang phase, when the
radiation dominates the background while the axion dominates the
fluctuations. In this case Eq. (\ref{22}) gives immediately:
\beq
\zeta_k= {1 \over 4}~{\r_\sg \da_\sg(k) \over \r_r} =
 \frac{1}{4\rho_{r}} \frac{\partial V}{\partial\sigma}  \chi_{k} =
\frac{1}{24} \frac{a^2}{{\cal H}^2} \frac{\partial V}{\partial\sigma}
\chi_{k} \propto a^4,
\label{221}
\eeq
where we have used the fact that, in the initial phase, $\sg$ is
approximately constant. Since  also $\Phi_{k}$ will behave like $a^4$, it
is easy to find its relation to $\zeta$ using, inside (\ref{21}), $\Phi' =
4 {\cal H} \Phi$ and ${\cal H}' = - {\cal H}^2$, with the result:
\beq
\Phi_{k} = - \frac{2}{7} \zeta_{k} =
 - \frac{1}{14}~{\r_\sg \da_\sg(k) \over \r_r}.
\label{222}
\eeq
In order to proceed further, two alternatives (already discussed
in the context of the background evolution) should now be separately
examined:

$(1)$ If $\sg_i < 1$, during the oscillating (but still 
radiation-dominated) phase, $\zeta$ can still  be obtained from Eq.
(\ref{22}), but now  $\r_\sg \sim \da \r_\sg \sim a^{-3}$, and $\zeta$ 
will evolve like $a \sim \eta \sim t^{1/2}$. Since $a$ changes by a factor
$(m/H_{\sg})^{1/2} \simeq (\sg_i)^{-2}$, we end up with a value of
$\zeta_{k}$ at $t_{\sg}$ given by:
\beq
 \zeta_k(t_\sg)  \sim  {\chi_k(t_i)\over \sg_i},
~~~~~~~~~~~~~~~~\sg_i \le 1 .
\label{161}
\eeq
On the other hand, using again Eq. (\ref{21}) and
the appropriate relations in the oscillating, radiation-dominated phase, 
we find $\Phi_k(t_\sg)  = - \zeta_k(t_\sg)/2$. 
In the final phase, dominated by an oscillating axion,
$\r_r$ is negligible, the (average) axion pressure is zero, and
(the average of) $\Phi_{k}$ is constant, as well as the average of
$\zeta_k$, which oscillates around a final amplitude of the same order
as $\zeta_k(t_\sg)$ given in eq. (\ref{161}). 
This implies, through Eqs.
(\ref{6}), (\ref{7}) and (\ref{12}),  $  \delta_{\sigma}(k) = - 2 \Phi_{k}
= -
(1/2) \delta_{r}(k)$, so that, from Eq. (\ref{21}) we are led to
\beq
\langle\zeta_k\rangle  = -{5\over 3} \langle \Phi_k \rangle =
{5\over 6} \langle \da_\sg (k) \rangle, 
\label{23}
\eeq
where $\langle ... \rangle$ refers to averages over one oscillation
period. We have checked the validity of this result  by an explicit
numerical integration (the same result has already been  presented in 
\cite{14}, using different notations). 

$(2)$ If $\sg_i > 1$, then  Eq. (\ref{22}) can still be used until $t_\sg
=1/m\sg_i$, where we find:
\beq
\zeta_k(t_\sg) =
 \frac{1}{4\r_{\sigma}} ~\frac{\partial V}{\partial\sigma}
 \chi_{k}  \sim  {\chi_k(t_i) \over \sg_i}.
\label{231}
\eeq
During the period of axion-dominated slow-roll inflation, Eq.  (\ref{22})
is still valid. However, since  $\r_r$  soon becomes
  subdominant with respect to $\r_{\sg} + p_{\sg}$,
it should be appreciated that at the end of the slow-roll
period the latter term is of order $m^2$, and  the resulting estimate
will thus be: 
\beq
\zeta_k(t_m) =
 {1 \over 4}~\frac{\partial V}{\partial\sigma}
{ \chi_{k} \over m^2} \sim  {\chi_k(t_i)  \sg_i},
~~~~~~~~~~~~~~~\sg_i>1. 
\label{232}
\eeq
Note that this formula is  in (qualitative) agreement with Eq.
(\ref{23}), 
if we use $\da \r_{\sg} \sim m^2 \chi_k(t_i)  \sg_i$ and
$\r_{\sg} \sim m^2$. No further amplification  is expected in the course
of the subsequent cosmological evolution. Similar expressions hold for
the amplitude of $\Phi_{k}$, related to $\zeta_k$ by Eq. (\ref{23}). 

It is amusing to observe that the  results (\ref{161}), (\ref{232}), which
determine the amplitude of the Bardeen potential in the oscillating
(axion-dominated) phase preceding the moment ($t = t_{d}$)
 at which the 
decay occurs,
can be summarized by an equation that holds in all cases, namely
\beq
\langle \Phi_k(t_d) \rangle =-\chi_k(t_i)f(\sg_i), ~~~~~~~~~~~
f(\sg_i)= \left(c_1 \sg_i+{c_2\over
\sg_i}+c_3\right),
\label{20}
\eeq
where $c_1,c_2,c_3$ are numerical coefficients of the  order of unity. 
A preliminary fit based on numerical and analytical integrations of
the perturbation equations gives $c_1=0.129,c_2=0.183,c_3=0.019$ (see
\cite{BGGV} for further details).  
The function $f(\sg_i)$ has the interesting feature that it is
approximately invariant under the transformation $\sg_i \ra \sg_i^{-1}$
and, as a consequence,  has a {\it minimal} value around $\sg_i =1 $, 
a result we shall use later on.

The generated spectrum of super-horizon curvature perturbations is
thus directly determined by the primordial spectrum of isocurvature
axion fluctuations $\chi_k$, according to Eqs. (\ref{161}) and (\ref{232}).
The axion fluctuations, on the other hand, are solutions
(with pre-big bang initial conditions) of Eq.
(\ref{9}) in the radiation era (no additional amplification is
expected, for super-horizon modes, in the axion-dominated phase),
computed  for negligible curvature perturbations
($\Phi=0=\Phi'$), evaluated  in the massive, non-relativistic limit
(where we are, eventually, in the oscillating regime) and outside the
horizon. The exact solution for $\chi_k$, normalized to a relativistic
spectrum of quantum fluctuations (amplified with the Bogoliubov
coefficient $c_k$) has already been  computed in \cite{8}. 
Setting $x= \eta\sqrt{2 \a}$, $\a=mH_1a_1^2$, $b=-k^2/2\a$,
it can be written in the form
\beq
\chi_k={c_k\over a} \left(k\over 2\a\right)^{1/2} y_2(b,x),
\label{24}
\eeq
where $y_2$ is the odd part of the parabolic cylinder functions
\cite{19}. Outside the horizon ($-b x^2 \ll 1$) and for
 non-relativistic modes ($-b \ll x^2$), the solution can be expanded, to
leading order, as $y_2 \sim x = \eta\sqrt{2 \a}$. By inserting a
generic power-law spectrum, with cut-off scale $k_1=H_1a_1$ and
spectral index $n$, i.e. $|c_k|=(k/k_1)^{(n-5)/2}$, we finally
obtain the generated spectrum of curvature perturbations:
\beq
k^3 \left| \Phi_k\right|^2 = f^2(\sg_i){k^3} \left|
\chi_k\right|^2
= f^2(\sg_i)\left(H_1\over  M_{\rm P}\right)^2
\left( k\over k_1\right)^{n-1}, ~~~~~~~~~~~~~~~k<k_1
\label{25}
\eeq
(we have absorbed into the definition of $k_1$ possible numerical
factors of order one connecting the cut-off scale to the string mass).

Note that we have re-inserted the appropriate Planck mass factors,
keeping  $\sg_i$ dimensionless. It may be useful to recall that the 
spectral index $n$ depends upon the pre big-bang dynamics \cite{6}, and
that for an isotropic $6$-dimensional subspace it can be written in the
form \cite{10} 
\begin{equation}
n = \frac{ 4 + 6 r^2 - 2 \sqrt{ 3 + 6 r^2}}{ 1 + 3 r^2},
\label{index}
\end{equation}
where $ r = (\dot{V}_{6} V_3)/(2 V_{6} \dot{V}_{3})$
  accounts for the relative rate of variation of the six-dimensional
  internal volume $V_6$ and of the ``external" 
(usual)  volume $V_3$. As already mentioned, the case
   of a flat spectrum (i.e. $n=1$)
   corresponds to $r=\pm 1$. Otherwise, $n$ increases 
monotonically with $r^2$ from the value  $ n = 4  - 2 
   \sqrt{3} \simeq 0.53$ when internal dimensions 
are static ($r=0$), to $n=2$ for the case of a static external 
   manifold ($r \to \infty $).

The result  (\ref{25}) is valid during the
axion-dominated phase, and has to be transferred to the phase of
standard evolution, by matching the (well-known \cite{17}) solution for
the Bardeen potential in the radiation era (subsequent to axion decay) 
to the solution prior to decay, which is in general oscillating. 
The matching of $\Phi$ and $\Phi'$, conventionally 
performed at the fixed scale $H=H_d$, shows that the constant
asymptotic value (\ref{20}) of super-horizon modes is preserved (to
leading order) by the decay process, modulo a random, mass-dependent
correction which typically takes the form $1+ \ep\sin (m/H_d)$, with
$\ep$ a numerical coefficient of order $1$, and $m/H_d \gg1$. Such a
random factor, however,  is a consequence of the sudden
approximation adopted to describe the decay process, and disappears
in a more realistic treatment in which the axion equation (\ref{3}) is
supplemented by the friction term $+\Gamma {\sigma'}/{a}$ (leading to
the term $+\Gamma {{\sigma'}^2}/{a^2}$ in the equation for
$\rho_\sigma$), and a corresponding antifriction term $-\Gamma
{{\sigma'}^2}/{a^2}$ in the radiation equation. The axion
fluctuations will follow the background and decay with a similar term,
$+\Gamma {\chi'}/{a}$, in the perturbation equation (\ref{9}). 

The previous analysis performed up to $t=t_d$ remains valid for the
modified equations, since for $\Gamma \ll H$  the decay terms are
negligible. We have checked with a numerical integration \cite{BGGV} 
that the decay process preserves the  value of the Bardeen
potential prior to decay, damping the residual oscillations; $\zeta$
itself follows the
same behaviour and is finally exactly a constant. When the axion has
completely decayed, and the Universe is again dominated by radiation, 
we can properly match the standard evolution of $\Phi$ in the radiation
phase to the constant asymptotic value of Eq. (\ref{20}). The expression
we obtain for the  (oscillating) Bardeen potential, valid until the epoch
of matter--radiation equality (denoted in the following by $\eta_{\rm
eq}$), can be written in the form 
\begin{equation}
\Phi_{k}(\eta) =-3\Phi_{k}(\eta_{d})  \biggl[ \frac{\cos{(
k c_s \eta)}}{(k c_s \eta)^2} - \frac{\sin{(k c_s\eta)}}{(k
c_{s}\eta)^3} \biggr] , ~~~~~~~~~\eta_d < \eta< \eta_{\rm eq}, 
\label{barfin}
\end{equation}
where $c_{s} = 1/\sqrt{3}$ and $\Phi_{k}(\eta_{d})$ is given in Eq.
(\ref{20}).

The above expression for the Bardeen potential  provides the initial
condition for the evolution of the CMB-temperature  fluctuations, 
and the formation of their oscillatory pattern. Standard results \cite{hs}
(see also \cite{sw}) imply that the  patterns of the CMB anisotropies
(and, in particular, the position of the first Doppler peak) are related
to the sum of two oscillating contributions, with a relative phase of
$\pi/2$. Denoting by $\eta_{\rm dec}$ the decoupling time, the first
contribution oscillates like
$A \cos{[k r_s(\eta_{\rm dec})]}$,  while the  second one
oscillates like $B \sin{[k r_{s}(\eta_{\rm dec})]}$,  where
$r_s(\eta_{\rm dec})$ is the sound-horizon at $\eta=\eta_{\rm dec}$. 
The value of $\Phi_{k}$ for  $\eta \ll \eta_{\rm eq} < \eta_{\rm
dec}$ determines, in particular,  the relative phase of oscillation of the
two terms. In our case, from Eq. (\ref{barfin}), $\Phi_{k}(\eta_i) = {\rm
constant}$ and $\Phi'_{k}(\eta_i) \simeq 0$, where
$\eta_d<\eta_i<\eta_{\rm eq}$, and $ k\eta_i \ll1$. This implies  $B =0$,
so that the temperature anisotropies $(\Da T/T)_k$ will oscillate  like
\cite{hs}  $ \Phi_{k}(\eta_i)\cos{[k r_s(\eta_{\rm dec})]}$, 
as is generally the  case for adiabatic fluctuations. The opposite
case,
$\Phi_{k} (\eta_i) \simeq 0$ and $\Phi_{k}'(\eta_i) =$ constant,
corresponds
instead to  isocurvature initial conditions \cite{KS}, producing a
peak structure that is clearly distinguishable from the adiabatic case and, at
present, observationally disfavoured. 

After checking that the above scenario leads to the
standard adiabatic mode, producing the observed  peak structure of
the CMB anisotropies, we still have  to discuss the possibility of a
correct large-scale normalization of the spectrum, compatible with
the COBE data. We start from the observation that 
the final amplitude of the super-horizon perturbations (\ref{25}), just
like the spectral slope, is not at all affected by the non-relativistic
corrections to the axion spectrum \cite{9},  in spite of the crucial role
played by the mass in the decay process (see also \cite{12}). The mass
dependence reappears, however, when computing the amplitude of the
spectrum at the present horizon scale $\om_0$, in order to impose the
corrected normalization to the quadrupole coefficient $C_2$ determined
by COBE, namely  \cite{Dur01}
\beq
C_2= 
\a^2_n f^2(\sg_i) 
\left(H_1\over M_{\rm P}\right)^2\left(\om_0\over
\om_1\right)^{n-1},  ~~~~~~~
\a_n^2 = {2^{n}\over 72}{\Ga (3-n) \Ga\left({3+n\over 2}\right)
\over \Ga^2\left({4-n\over 2}\right) \Ga\left({9-n\over 2}\right)},
\label{alpha}
\eeq
where \cite{20} $C_2=(1.9\pm 0.23)\times 10^{-10}$. 

The present value of the cut-off
frequency, $\om_1(t_0)=H_1a_1/a_0$, depends in fact on the kinematics
as well as  on the duration of the axion-dominated phase (and thus on
the axion mass), as follows: \bea
\om_1(t_0)  &=& H_1
\left(a_1\over a_\sg\right)_{\rm rad} \left(a_\sg\over
a_d\right)_{\rm mat} \left(a_d\over a_{\rm eq}\right)_{\rm rad}
\left(a_{\rm eq}\over a_0\right)_{\rm
mat}, ~~~~~~~~~~~~~~~~~ \sg_i < 1,
\label{28}\\
&=&H_1\left(a_1\over a_\sg\right)_{\rm
rad} \left(a_\sg\over a_{\rm
osc}\right)_{\rm inf} \left(a_{\rm
osc}\over a_d\right)_{\rm
mat} \left(a_d\over a_{\rm eq}\right)_{\rm
rad} \left(a_{\rm eq}\over a_0\right)_{\rm
mat}, ~~~ \sg_i >1. 
\label{29}
\eea
Using $H_0 \simeq 10^{-6} H_{\rm eq} \simeq 10^{-61}
M_{\rm P}$ we find 
\bea
{\om_0\over \om_1} &\simeq & 10^{-29}
\left(H_1\over  M_{\rm P}\right)^{-1/2}
\left(m\over \sg_i^2 M_{\rm P}\right)^{-
1/3}, ~~~~~~~~~~~~~~~~~\sg_i < 1,
\label{30}\\
&\simeq & 10^{-29}
\left(\sg_i H_1\over  M_{\rm P}\right)^{-
1/2} \left(m\over \ M_{\rm P}\right)^{-
1/3}Z_\sg, ~~~~~~~~~~~~~~\sg_i > 1,
\label{31}
\eea
where $Z_\sg= (a_{\rm osc}/a_\sg)$ denotes the
amplification of the scale factor during the phase of axion-dominated,
slow-roll inflation.
The COBE normalization thus imposes
\bea
&&
c_2^2\a^2_n \sg_i^{2(n-4)/3}
\left(H_1\over M_{\rm P}\right)^{(5-n)/2}
\left(m\over M_{\rm P}\right)^{-(n-1)/3} 10^{-
29(n-1)} \simeq 10^{-10}, ~~~~~~~~\sg_i < 1,
\label{32}\\
&&
c_1^2\a^2_n Z_\sg^{n-1} \sg_i^{(5-n)/2}
\left(H_1\over M_{\rm P}\right)^{(5-n)/2}
\left(m\over M_{\rm P}\right)^{-(n-1)/3} 10^{-
29(n-1)} \simeq 10^{-10}, ~~~\sg_i >1.
\label{33}
\eea
We can notice, as a side remark, that  the
contribution of the gradients appearing in Eqs. (\ref{6})--(\ref{9})
follows the same hierarchy of scales as provided by Eqs. (\ref{30}), 
(\ref{31}) and this is the reason why, ultimately, the contribution of
the
gradients can be neglected as far as the evolution of large-scale
modes is concerned.

The condition (\ref{32}) is to be combined with the constraint
(\ref{13}), the condition (\ref{33}) with the constraint (\ref{18}),
which are required for the consistency of the corresponding classes of
background evolution. Also, both conditions are to be intersected
with the experimentally allowed range of the spectral index.
Finally, in the case $\sg_i >1$ we are also implicitly 
assuming that the axion-driven inflation is short enough to avoid a
possible contribution to $C_2$ arising from the metric fluctuations
directly amplified from the vacuum, during the phase of axionic
inflation. This requires that the smallest amplified frequency mode
$\om_\sg$, crossing the horizon at the beginning of inflation, be today
 still larger than the present horizon scale $\om_0$. This imposes
the condition  $\om_\sg(t_0) =H_\sg (a_\sg/a_0)>\om_0$, namely
\beq
Z_\sg \laq 10^{29} \sg_i \left(m\over M_{\rm P}\right)^{5/6},
\label{34}
\eeq
to be added to the constraint (\ref{18}) for $\sg_i >1$. It turns
out, however, that this condition is always automatically satisfied for
the range of spectral indices we are interested in (in particular, for 
$n \leq 1.7$). 

The allowed range of parameters compatible with all constraints is
rather strongly sensitive to the values of the pre-big bang
inflation scale $H_1$. In the context of minimal models of pre-big
bang inflation \cite{1} we have $H_1 \sim M_s$, and a flat spectrum
($n=1$) is inconsistent with the normalization (\ref{32}),
(\ref{33}). A growing (``blue") spectrum is instead allowed, and by
setting for instance $c_2\a_n H_1/ M_{\rm P}=10^{-2}$, using (as a
reference value) the upper bound \cite{21} $n<1.4$, and considering the
case $\sg_i \leq 1$, we find a wide range of allowed axion masses, but a
rather narrow range of allowed values  for $\sg_i$, namely $1  \gaq
\sg_i \gaq 10^{-5/2}$, and of allowed values for the spectral index, $n
\simeq  1.22$--$1.4$. In the case $\sg_i >1$ the results are 
complementary for the spectral index, but there are much more
stringent bounds for $\sg_i$, because the inflationary redshift factor
$Z_\sg$ grows exponentially with $\sg_i^2$, in such a way that the COBE
normalization (\ref{33}) cannot be satisfied, unless the upper value of
$\sg_i$ is strongly bounded. This means that the apparent symmetry
between the $\sigma_{i} <1$ and
the $\sigma_{i} > 1$ cases is broken by the requirement of the
CMB normalization,  which forbids too large values of $\sigma_{i}$.

The allowed region may be further extended if the inflation scale $H_1$
is lowered, and a flat ($n=1$) spectrum may become possible if $c_2
\a_n H_1 \laq 10^{-5}M_{\rm P} \sg_i$, for $\sg_i<1$, and if $c_1\a_n
H_1 \laq 10^{-5}M_{\rm P} /\sg_i$, for $\sg_i>1$ (see Eqs. (\ref{32}) and
(\ref{33})). This possibility could arise in a recently proposed
framework \cite{strong} according to which, at strong bare coupling
$e^{\phi}$,
 loop effects renormalize downwards the ratio $M_s/M_{\rm P}$
 and allow $M_s$ to approach the unification scale. In addition,  a flat
spectrum may be allowed even keeping pre-big bang inflation at a
high-curvature  scale, provided the relativistic
branch of the primordial axion fluctuations is characterized by a
frequency-dependent slope, which is flat enough at low frequency (to
agree with large-scale observations) and much steeper at high
frequencies (to match the string normalization at the end-point of the
spectrum).
                                  
A typical example of such a spectrum can be parametrized by a
Bogoliubov coefficient with a break at the intermediate scale
$k_s$, \bea
|c_k|^2 &=& \left(k\over k_1\right)^{n- 5 + \delta},
~~~~~~~~~~~~~~~~~~~~k_s<k<k_1,
\nonumber\\
&=& \left(k_s\over k_1\right)^{n- 5 + \delta}
\left(k\over k_s\right)^{n-5},
~~~~~~~~~~~~ ~~k<k_s,
\label{35}
\eea
where $\delta > 0 $ parametrizes the slope of the break at high
frequency. Examples of realistic pre-big bang backgrounds producing
such a spectrum of axion fluctuations have been already presented in
\cite{9}. 
Furthermore, a steeper axion spectrum at high frequency could also
emerge if the exit from pre-big bang inflation occurred at relatively
strong bare coupling, where various quantities may become
dilaton-independent as argued in \cite{strong}, and  the renormalized
axion pump field should approach the canonical pump field of metric
perturbations. Quite independently of the effective mechanism, 
it is clear that the steeper and/or the longer the
high-frequency branch of the spectrum, the larger the suppression
at low-frequency scales, and the easier the matching of the
amplitude to the measured anisotropies (in spite of possible
$\sg_i$-dependent enhancements).

Using the generalized input (\ref{35}) for the spectrum of $\chi_k$,
the amplitude of the low-frequency ($k<k_s$) Bardeen spectrum
(\ref{25}) is to be multiplied by the suppression factor
$\Da=(k_s/k_1)^{\delta} \ll 1$,
and the normalization condition at the COBE
scale becomes
\bea
&&
\a^2_n c_2^2 \left(H_1\over \sg_i M_{\rm P}\right)^{2}
\left(\om_0\over \om_1\right)^{n-1}
 \simeq C_2\Da^{-1},
~~~~~~~~~~\sg_i <1,
\label{36}\\
&&
\a^2_n c_1^2 \left(\sg_i H_1\over M_{\rm
P}\right)^{2} \left(\om_0\over
\om_1\right)^{n-1}
 \simeq C_2\Da^{-1},
~~~~~~~~~~~\sg_i >1.
\label{37}
\eea
A strictly flat spectrum is now possible, even for $\a_1 H_1=\a_1
M_s\simeq 10^{- 2}M_{\rm P}$, provided
\beq
\Da \left(c_1^2\sg_i^2
+ c_2^2\sg_i^{-2} \right) \laq 10^{-6}. 
\label{38}
\eeq
It thus becomes  possible, in this context, to satisfy the stringent
limits imposed by the most recent analyses of the peak and dip
structure of the spectrum at small scales \cite{22}, which imply $0.87
\leq n \leq 1.06$ (see also \cite{22a}). 

In order to illustrate this
possibility, let us specify further Eq. (\ref{35}) by identifying $k_s$
with the scale $k_{\rm eq}$ of  matter--radiation equivalence, in
such a way that $n$ will denote the value of the axion spectral index
for the scales relevant to CMB anisotropies, while $n+\delta$ 
provides the (average) axion spectral
index in the remaining range of scales, up to the cutoff $k_1$. Then, 
after imposing the COBE normalization condition 
$\a_n^2 f^2(\sg_i)(H_1/M_{\rm P})^2(\om_0/\om_1)^2=C_2$, we plot in Fig.
1  curves corresponding to some given values of the ratio $H_1/M_{\rm P} 
\sim M_s/M_{\rm P}$.  We have done this choosing the values
$\sg_i=1$ and $m = 10^{10}$ GeV, but for $n$ around $1$ the curves are
very stable, even if we change $m$ by many orders of magnitude,
provided we stay at $\sg_i$ of order $1$ (i.e. near the minimum of $f$).
A look at the figure shows immediately that the phenomenologically
allowed range for $n$ is theoretically consistent even for
$M_s/M_{\rm P}$ as large as $0.1$, provided we allow for a small break 
in the spectrum, $\delta \simeq 0.2$. Conversely, we can allow having
no break at all in the spectrum ($\delta = 0$), if we are willing to take
$M_s/M_{\rm P} \sim 10^{-3}$, i.e. a string mass  close to the GUT scale.

\begin{figure}[t]
\centerline{\epsfig{file=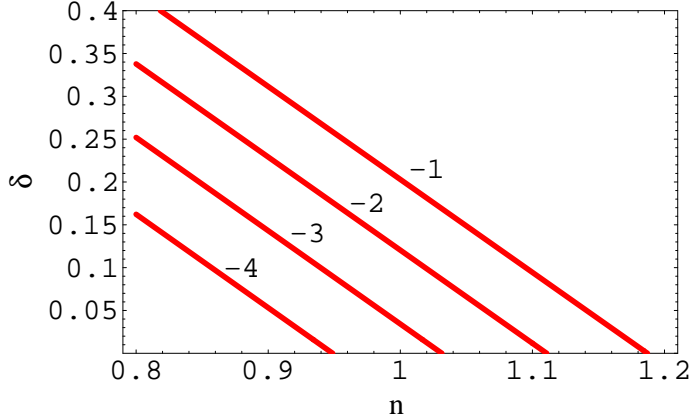,width=92mm}}
\vskip 3mm
\caption{\sl Plot of the COBE normalization condition for $\sg_i=1$,
$f(1)=0.33$, $m=10^{10}$ GeV, $k_1/k_s=k_1/k_{\rm eq}\simeq 10^{27}
(H_1/M_{\rm P})^{1/2}(m/M_{\rm P})^{1/3}$, and  for various values of
the inflation scale $H_1$. The four curves correspond, from left to right
respectively, to $\log (H_1/M_{\rm P})=-4,-3,-2,-1$. }  
\label{f1} 
\end{figure}

We conclude that,  in the context of the pre-big bang scenario, a
``curvaton" model  based on the Kalb--Ramond axion is able to
produce the adiabatic curvature perturbation needed to explain the
observed large-scale anisotropies. The simplest, minimal model of
pre-big bang inflation seems to prefer blue spectra. A strictly
scale-invariant (or even slightly red, $n<1$) spectrum is not excluded 
but requires, for normalization purposes,
non-minimal models of pre-big bang evolution leading to axion
fluctuations with a sufficiently steep slope at high frequencies.

\end{document}